\documentclass[letter]{jpsj2} 
\title{
Coexistence of Supercondcutivity and Magnetism in 
LaFeAs(O$_{0.94}$F$_{0.06}$) Probed by Muon Spin Relaxation
}

\author{
Soshi \textsc{Takeshita}$^{1}$\thanks{E-mail address : soshi@post.kek.jp},
Ryosuke \textsc{Kadono}$^{1,2}$,
Masatoshi \textsc{Hiraishi}$^{2}$,\\
Masanori \textsc{Miyazaki}$^{2}$,
Akihiro \textsc{Koda}$^{1,2}$,
Yoichi \textsc{Kamihara}$^{3}$
and
Hideo \textsc{Hosono}$^{3,4,5}$
}

\inst{
$^{1}$Institute of Materials Structure Science, High Energy Accelerator
Research Organization (KEK), 1-1 Oho, Tsukuba, Ibaraki 305-0801, Japan\\
$^{2}$Department of Materials Structure Science, School of High Energy
Accelerator Science, The Graduate University for Advanced Studies,
1-1 Oho, Tsukuba, Ibaraki 305-0801, Japan\\
$^{3}$ERATO-SORST, JST, in Frontier Research Center, Tokyo Institute of
Technology, 4259 Nagatsuta, Midori-ku, Yokohama, Kanagawa 226-8503,
Japan\\
$^{4}$Frontier Research Center, Tokyo Institute of Technology, 4259
Nagatsuta, Midori-ku, Yokohama, Kanagawa 226-8503, Japan\\
$^{5}$Materials and Structures Laboratory, Tokyo Institute of Technology,
4259 Nagatsuta, Midori-ku, Yokohama, Kanagawa 226-8503, Japan\\
}

\abst{
The presence of macroscopic phase separation into superconducting and 
spin-glass-like magnetic phases in LaFeAs(O$_{1-x}$F$_x$) is demonstrated 
by muon spin rotation measurement in a sample near the phase boundary 
($x=0.06$).  Both magnetism and superconductivity develop simultaneously 
below a common critical temperature, $T_{\rm m}\simeq T_{\rm c}\simeq18$ K.  
This remarkable accordance strongly suggests that 
the electronic correlations leading to these two competing ground states
share a common origin.
}

\kword{oxypnictide superconductor, muon spin rotation, electronic correlation, 
magnetism}

\begin{document}
\maketitle


The recent discovery of the oxypnictide superconductor
LaFeAsO$_{1-x}$F$_x$ (LFAO-F) with a critical temperature ($T_{\rm c}$)
of 26~K \cite{R_Kamihara} and the successful revelation of much increased
$T_{\rm c}$ upon the substitution of La for other rare-earth elements
(such as Sm, leading to $\sim43$~K \cite{R_SmFAOF}) and the application
of pressure for LFAO-F ($\sim43$~K \cite{R_Pressure}) have triggered
broad interest in the mechanism yielding a relatively high $T_{\rm c}$
in this new class of compounds.  They have a layered structure like
high-$T_{\rm c}$ cuprates, where the dopant and conducting layers are so
separated that the doped carriers (electrons introduced by the
substitution of O$^{2-}$ with F$^{-}$ in the La$_2$O$_2$ layers) move
within the layers consisting of strongly bonded Fe and As atoms.  They
exhibit another qualitative similarity to cuprates in that
superconductivity occurs upon carrier doping of pristine compounds that
exhibit magnetism. \cite{Cruz:08} Some preliminary results of the muon
spin rotation/relaxation ($\mu$SR) experiment on a variety of
oxypnictide superconductors showed that the superfluid density $n_s$
falls on the empirical line on the $n_s$ vs $T_{\rm c}$ diagram
observed for the {\sl underdoped} cuprates, \cite{Luetkens:08,Carlo:08}
from which possibility of the common mechanism of superconductivity is
argued between oxypnictides and cuprates.

However, in terms of the doping phase diagram, there are certain
differences between these two systems, e.g., (i) $T_{\rm c}$ ($>0$ for
$0.4<x<0.12$) does not vary much with $x$ \cite{R_Kamihara} as in
cuprates known as ``bell-shaped'' and (ii) the magnetic (spin density
wave, SDW) phase shares a boundary with the superconducting phase near
$x\simeq0.04$.\cite{R_Kamihara,Luetkens:08-2} The insensitivity of
$T_{\rm c}$ to $x$ is reasonably understood from the conventional BCS
theory where condensation energy is predicted to be independent of
carrier concentration.  The close relationship of magnetism and
superconductivity suggests that a detailed investigation of how these
two phases coexist (and compete) near the phase boundary will provide
important clues to elucidating the paring mechanism. Among various
techniques, $\mu$SR has a great advantage in that it can be applied in
systems consisting of spatially inhomogeneous multiple phases, providing
information on respective phases according to their fractional yield.
Our $\mu$SR measurement in the LFAO-F sample with $x=0.06$ ($T_{\rm
c}\simeq18$ K) reveals that these two phases indeed coexist in the form of
macroscopic phase separation, and more interestingly, that a spin
glass-like magnetic phase develops in conjunction with superconductivity
in the paramagnetic phase.  This accordance strongly suggests a common
origin of the electronic correlation leading to these two competing
phases.

\par

\begin{figure}[tp]
\begin{center}
\includegraphics[width=0.6\textwidth,clip]{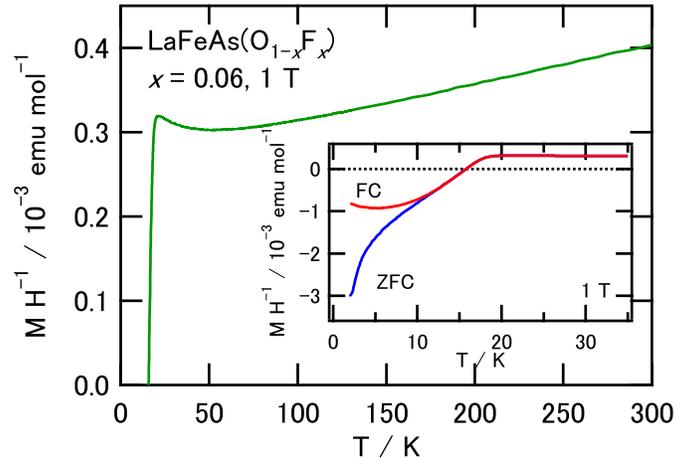}
\caption{(Color online)
Magnetic susceptibility of LaFeAsO$_{1-x}$F$_x$ with $x=0.06$
for the sample used for $\mu$SR measurement. Inset shows a 
reduced view of the region below 35 K.}
\label{dcChi}
\end{center}
\end{figure}

\begin{figure}[tp]
\begin{center}
\includegraphics[width=0.5\textwidth,clip]{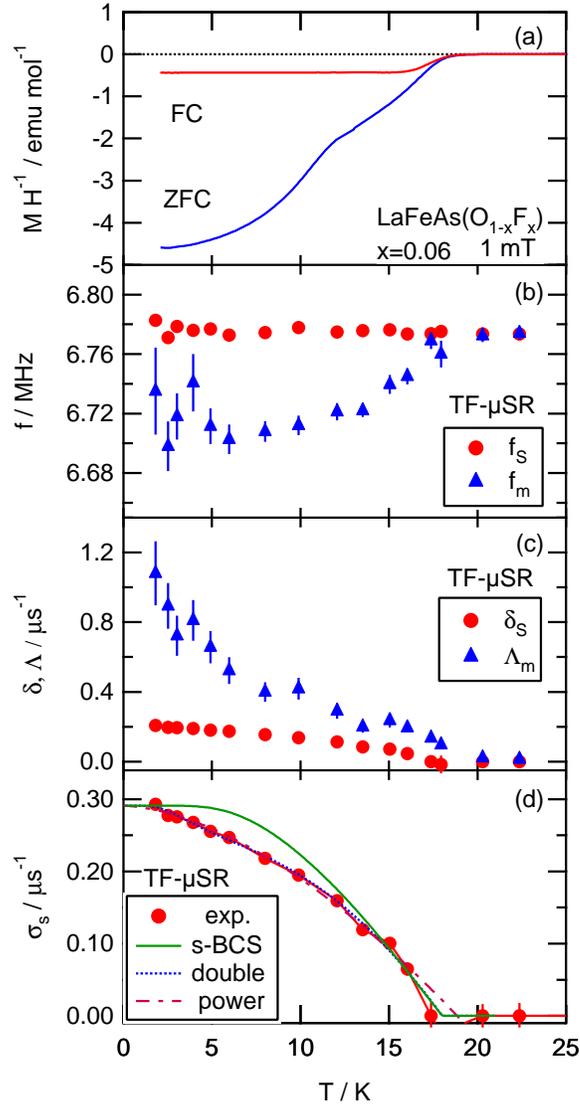}
\caption{(Color online)
Temperature dependence of dc magnetic susceptibility measured at 1mT
 (a), and that of physical parameters deduced by analyzing TF-$\mu$SR
 spectra in superconducting ($f_{\rm s}$, $\delta_{\rm s}$) and magnetic
 ($f_{\rm m}$, $\Lambda_{\rm m}$) phases (b-c), and of $\sigma_{\rm s}$
 ($=\sqrt{2}\delta_{\rm s}$) proportional to superfluid density
 (d). Curves in (d) are fits by models described in the text.
}
\label{G_TFmulti}
\end{center}
\end{figure}

Although the oxypnictide with rare-earth ($R$) substitution
$R$FeAsO$_{1-x}$F$_x$ exhibits higher $T_{\rm c}$ than that of LFAO-F,
strong random fields from rare-earth ions preclude a detailed study of
the ground state using sensitive magnetic probes like $\mu$SR.
Therefore, we chose the original LFAO-F system for our $\mu$SR study.
The target concentration of LaFeAsO$_{1-x}$F$_x$ is set near the phase
boundary, $x=0.06$, for which a polycrystalline sample was synthesized
by solid state reaction. The detailed procedure for sample preparation
is described in an earlier report \cite{R_Kamihara}.  The sample was
confirmed to be mostly of single phase using X-ray diffraction
analysis. Of two possible impurity phases, namely, LaOF and FeAs, only the
latter exhibits a magnetic (helical) order with $T_N\simeq77$
K.\cite{Selte:72} As shown in Fig.~\ref{dcChi}, magnetic susceptibility
exhibits no trace of FeAs phase or local magnetic impurities except
below $\sim50$ K where a small upturn is observed.  The susceptibility
at a lower field [shown in Fig.~\ref{G_TFmulti}(a)] provides evidence of
bulk superconductivity with $T_{\rm c}\sim18$~K from the onset of
diamagnetism. Conventional $\mu$SR measurement was performed using the
LAMPF spectrometer installed on the M15 beamline of TRIUMF, Canada.
During the measurement under a zero field (ZF), residual magnetic field
at the sample position was reduced below $10^{-6}$~T with the initial
muon spin direction parallel to the muon beam direction
[$\vec{P}_\mu(0)\parallel \hat{z}$].  For longitudinal field (LF)
measurement, a magnetic field was applied parallel to $\vec{P}_\mu(0)$.
Time-dependent muon polarization [$G_z(t)=\hat{z}\cdot \vec{P}_\mu(t)$]
was monitored by measuring decay-positron asymmetry along the
$\hat{z}$-axis.  Transverse field (TF) condition was realized by
rotating the initial muon polarization so that $\vec{P}_\mu(0)\parallel
\hat{x}$, where the asymmetry was monitored along the $\hat{x}$-axis to
obtain $G_x(t)=\hat{x}\cdot \vec{P}_\mu(t)$.  All the measurements under
a magnetic field were made by cooling the sample to the target
temperature after the field equilibrated.

\par

\begin{figure}[tp]
\begin{center}
\includegraphics[width=0.6\textwidth,clip]{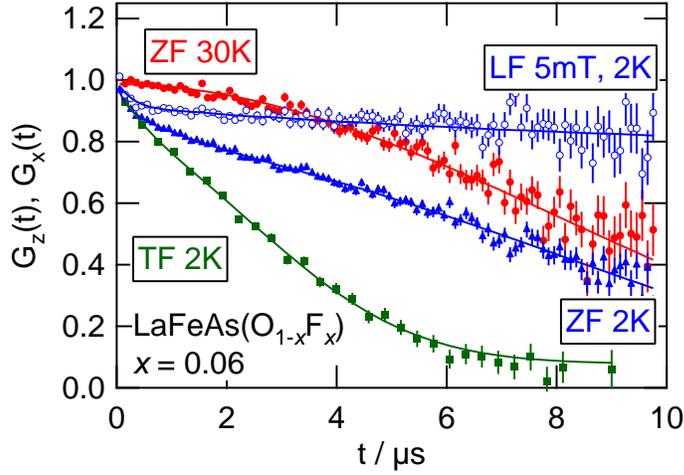}
\caption{(Color online)
$\mu$SR time spectra observed in LaFeAs(O$_{1-x}$F$_x$) with
$x=0.06$ at 2 K under a longitudinal field (LF), a zero field (ZF), and
a transverse field (TF), and that under ZF at 30 K.
The spectrum under TF is plotted on a rotating reference
frame of 6.78 MHz to extract the envelop function. }
\label{G_TSAll}
\end{center}
\end{figure}

ZF-$\mu$SR is the most sensitive technique for examining magnetism in
any form, where the development of local magnetic moments leads to
either the spontaneous oscillation (for long-range order) or exponential
damping (inhomogeneous local magnetism) of $G_z(t)$.  Figure
\ref{G_TSAll} shows examples of ZF-$\mu$SR time spectra collected at 2
and 30~K.  The spectrum at 30~K ($>T_{\rm c}$) exhibits a Gaussian-like
relaxation due to weak random local fields from nuclear magnetic
moments, indicating that the entire sample is in the paramagnetic state.
Meanwhile, the spectrum at 2~K is split into two components, one that
exhibits a steep relaxation and the other that remains to show
Gaussian-like relaxation. This indicates that there is a finite fraction
of implanted muons that sense hyperfine fields from local electronic
moments.  The absence of oscillatory signal implies that the hyperfine
field is highly inhomogeneous, so that the local magnetism is
characterized by strong randomness or spin fluctuation.  The fractional
yield of the component showing steep relaxation is as large as 25\% (see
below), which is hardly attributed to impurity and therefore implies
that the sample exhibits a macroscopic phase separation into two phases.

\par

The magnitude of the hyperfine field and that of spin fluctuation are
evaluated by observing the response of the $\mu$SR spectrum to a
longitudinal field (LF).  It is shown in Fig.~\ref{G_TSAll} that the
relaxation in the paramagnetic component is quenched by applying a weak
magnetic field (LF=5 mT), which is perfectly explained by the
suppression of static nuclear dipolar fields ($<10^0$ mT).  Meanwhile,
the faster relaxation (seen for $0<t<1$ $\mu$s) due to the magnetic
phase is recovered only gradually over a field range of $10^{1\sim2}$
mT, and there still remains a slow relaxation even at the highest field
of 60 mT.  This residual depolarization under LF is a clear sign that
local spins are slowly fluctuating, leading to the spin-lattice
relaxation of $\vec{P}_\mu(t)$.  Such quasi-two-step relaxation is also
observed in dilute spin-glass systems,\cite{Uemura:81} which is
understood as a distribution of spin correlation time.  A detailed
analysis is made considering that these two components coming from the
magnetic phase (see below).

\par

Under a transverse field, implanted muons experience an inhomogeneity of
the field [$B_z({\bf r})$] due to flux line lattice formation below
$T_{\rm c}$ that leads to relaxation, in addition to those observed
under a zero field.  The TF-$\mu$SR time spectrum in Fig.~\ref{G_TSAll}
(envelop part of the oscillation) obtained under a field of 50 mT
exhibits complete depolarization at 2 K, indicating that the entire
volume of the paramagnetic phase falls into the superconducting state.
The rapidly relaxing component observed under ZF is also visible
(although the coarse binning of the spectra for extracting the envelop
part makes it slightly obscure), indicating that the corresponding part
of the sample remains magnetic.

\par

Considering the presence of the magnetic phase besides the paramagnetic
(=superconducting below $T_{\rm c}$) phase, we take special precaution
to analyze both TF and ZF/LF $\mu$SR spectra in a consistent manner.
For the determination of physical parameters describing the behavior of
signals from the magnetic phase, we first analyze ZF/LF spectra at 2~K
using the $\chi$-square minimization fit with the relaxation function
\begin{equation}
\label{E_LF}
G_z(t)=[w_1 + \sum^3_{i=2}w_i 
 \exp{(-\Lambda_i t)}]\cdot G_{\rm KT}(\delta_{\rm N}:t),
\end{equation}
where $G_{\rm KT}(\delta_{\rm N}:t)$ is the Kubo-Toyabe relaxation
function for describing the Gaussian damping due to random local fields
from nuclear moments (with $\delta_{\rm N}$ being the linewidth)
\cite{R_Hayano}, $w_1$ is the fractional yield for the paramagnetic
phase, $w_2$ and $w_3$ are those for the magnetic phase ($\sum w_i=1$)
with $\Lambda_2$ and $\Lambda_3$ being the corresponding relaxation rate
described by the Redfield relation
\begin{equation}
 \Lambda_{i} = \frac{2\delta_i^2 \nu_i}{\nu_i^2 + \omega_\mu^2}\label{lmdmg}
 \ \ \ (i=2,3),
\end{equation}
where $\omega_\mu=\gamma_\mu H_{\rm LF}$, $\gamma_\mu$ is the muon
gyromagnetic ratio ($=2\pi\times135.53$ MHz/T), $H_{\rm LF}$ is the
longitudinal field, $\delta_2$ and $\delta_3$ are the means of the
hyperfine fields exerting on muons from local electronic moments, and
$\nu_2$ and $\nu_3$ are the fluctuation rates of the hyperfine field.
The solid curves in Fig.~\ref{G_TSAll} show the result of analysis where
all the spectra at different fields (only ZF and LF=5mT are shown here)
are fitted simultaneously using eqs.~(\ref{E_LF}) and (\ref{lmdmg}) with
common parameter values (except $\omega_\mu$ that is fixed to the
respective value for each spectrum), which show excellent agreement with
all the spectra. The deduced parameters are as follows: $w_1=0.754(9)$,
$w_2=0.165(9)$, $w_3=0.081(4)$, $\delta_2=0.71(5)$ $\mu$s$^{-1}$,
$\delta_3=3.9(3)$ $\mu$s$^{-1}$, $\nu_2=1.7(2)$ $\mu$s$^{-1}$, and
$\nu_3=4(1)$ $\mu$s$^{-1}$.  Although the depolarization in the magnetic
phase is approximately represented by two components with different
hyperfine couplings ($\delta_i$), the fluctuation rates ($\nu_i$) are
close to each other (10$^7$ s$^{-1}$ at 2 K), suggesting that the
randomness is primarily due to the distribution in the size of local
moments (or in their distances to muons).  Since no impurity phase with
a fraction as large as 25\% is detected by X-ray diffraction analysis,
it is concluded that this magnetic phase is intrinsic.


\par

In the analysis of temperature-dependent TF spectra, we used the
relaxation function
\begin{eqnarray}
\label{E_TFG}
 G_x(t)&=&\exp(-\frac{1}{2}\delta^2_{\rm N}t^2)
 [ w_1\exp(-\delta^2_{\rm s}t^2)\cos{(2\pi f_{\rm s} t+\phi)}\nonumber\\
  &+& (w_2+w_3)\exp{(-\Lambda_{\rm m} t)}\cos{(2\pi f_{\rm m}t+\phi)}],
\end{eqnarray}
where $w_i$ and $\delta_{\rm N}$ are fixed to the values obtained by
analyzing ZF/LF-$\mu$SR spectra.  The first component in the above
equation represents the contribution of flux line lattice formation in
the superconducting phase, where $\delta_{\rm s}$ corresponds to the
linewidth $\sigma_{\rm s}=\sqrt{2}\delta_{\rm s}=\gamma_\mu\langle
(B({\bf r})-B_0)^2\rangle^{1/2}$ [with $B_0$ being the mean $B({\bf
r})$], while the second term represents the relaxation in the magnetic
phase. Here, the relaxation rate for the latter is represented by a
single value $\Lambda_{\rm m}$ (instead of $\Lambda_{2,3}$), as it turns
out that the two components observed under LF are hardly discernible in
TF-$\mu$SR spectra.  [This does not affect the result of the analysis,
because the amplitude is fixed to $w_2+w_3$ so that $\Lambda_{\rm m}$
may represent a mean $\simeq(w_2\Lambda_2+w_3\Lambda_3)/(w_2+w_3)$.]
The fit analysis using the above form indicates that all the spectra are
perfectly reproduced while the partial asymmetry is fixed to the value
determined from ZF-$\mu$SR spectra. This strengthens the presumption
that the paramagnetic phase becomes superconducting below $T_{\rm c}$.
The result of analysis is summarized in Fig.~\ref{G_TFmulti}, together
with the result of dc magnetization measured in the sample from the same
batch as that used for $\mu$SR.

 \par

It is interesting to note in Fig.~\ref{G_TFmulti}(b) that, although the
central frequency in the superconducting phase ($f_{\rm s}$) does not
show much change below $T_{\rm c}\simeq18$ K probably owing to a large
magnetic penetration depth (it is indeed large, see below), that in the
magnetic phase ($f_{\rm m}$) exhibits a clear shift in the negative
direction below $T_{\rm m}\simeq T_{\rm c}$. The magnitude of the shift
is as large as $\sim1$\% and thus is readily identified despite a
relatively low external field of 50 mT.  As shown in
Fig.~\ref{G_TFmulti}(c), the relaxation rate in the magnetic phase
($\Lambda_{\rm m}$) also develops below $T_{\rm c}$ in accordance with
the frequency shift, demonstrating that a spin-glass-like magnetism sets
in below $T_{\rm c}$.  Here, we note that the development of magnetic
phase is already evident in the ZF/LF-$\mu$SR spectra, and results are
fully consistent with each other.  The onset of superconductivity below
$T_{\rm c}$ is also confirmed by an increase in $\delta_{\rm s}$, as
observed in Fig.~\ref{G_TFmulti}(c).  This remarkable accordance of
onset temperature between magnetism and superconductivity strongly
suggests that there is an intrinsic relationship between the
superconducting and magnetic phases that leads to a common
characteristic temperature.

\par

The temperature dependence of $\sigma_{\rm s}$ in
Fig.~\ref{G_TFmulti}(d) is compared with theoretical predictions for a
variety of models with different order parameters. The weak-coupling BCS
model ($s$-wave, single gap) apparently fails to reproduce the present
data, as they exhibit a tendency to vary with temperature over the
region $T/T_{\rm c}<0.4$.  Although a fit using a two-gap model
\cite{R_TwoGap} shown by a dotted line seems to exhibit reasonable
agreement with data, the deduced gap parameters ($2\Delta_i/k_{\rm
B}T_{\rm c}$) are largely inconsistent with the prediction of the
weak-coupling BCS model (see Table \ref{T_SgmPara}).  These observations
suggest that the superconducting order parameter is not described by a
s-wave symmetry with a single gap.  When a power law, $\sigma_{\rm s} =
\sigma_0 [1-(T/T_{\rm c})^\beta]$, is used in fitting the data, we
obtain a curve shown by the broken line in Fig.~\ref{G_TFmulti}(d) with
an exponent $\beta\simeq2$.  This is in good agreement with the case of
$d$-wave symmetry at the dirty limit.

\begin{table}[bp]
\begin{center}
\caption{Parameters for defining the lines in Fig.~\ref{G_TFmulti}(d).}
\begin{tabular}{cc|cc}
\hline
Two-gap & & Power law& \\
 \hline
 $T_{\rm c}$ (K)  &  18.0(5) & $T_{\rm c} (K)$    &  18.9(4)  \\
 $\sigma(0)$ ($\mu {\rm s}^{-1}$) & 0.291(5) 
 & $\sigma(0)$ ($\mu {\rm s}^{-1})$ & 0.291(4)\\
 $w$                    &  0.73(6) & $\beta$    &  1.7(1) \\
 $2\Delta_1/k_{\rm B}T_{\rm c}$     &  2.6(3)  \\
 $2\Delta_2/k_{\rm B}T_{\rm c}$     &  1.1(3) \\
 \hline
\end{tabular}
\label{T_SgmPara}
\end{center}
\end{table}

\par

In the limit of extreme type II superconductors [i.e.,
$\lambda/\xi\gg1$, where $\lambda$ is the effective London penetration
depth and $\xi=\sqrt{\Phi_0/(2\pi H_{\rm c2})}$ is the Ginzburg-Landau
coherence length, $\Phi_0$ is the flux quantum, and $H_{\rm c2}$ is the
upper critical field], $\sigma_{\rm s}$ is determined by $\lambda$ using
the relation \cite{R_Brandt} $ \sigma_{\rm s}/\gamma_\mu=
2.74\times10^{-2}(1-h)\left[1+3.9(1-h)^2\right]^{1/2}\Phi_0\lambda^{-2}$,
where $h=H_{\rm TF}/H_{\rm c2}$ and $H_{\rm TF}$ is the magnitude of
external field.  From $\sigma_{\rm s}$ extrapolated to $T=0$ and taking
$H_{\rm c2}\simeq50$~T (ref.~\cite{R_Hc2}), we obtain
$\lambda$=595(3)~nm.  Because of the large anisotropy expected from the
layered structure of this compound, $\lambda$ in the polycrystalline
sample would be predominantly determined by in-plane penetration depth
($\lambda_{\rm ab}$).  Using the formula $\lambda=1.31\lambda_{\rm ab}$
for such a situation \cite{R_Fesenko}, we obtain $\lambda_{\rm
ab}$=454(2)~nm.  This value coincides with that expected from the
aforementioned empirical relation between $\lambda^{-2}_{\rm ab}$
superconductors \cite{R_Uemura,Carlo:08}.  However, this may not be
uniquely attributed to the superfluid density because $\lambda$ depends
not only $n_{\rm s}$ but also on the effective mass, $ \sigma_{\rm
s}\propto\lambda^{-2}= n_{\rm s}e^2/m^*c^2$.

\par

Finally, we discuss the feature of the spin glass-like phase.  Assuming
that the local moments are those of off-stoichiometric iron atoms with a
moment size close to that in the SDW phase ($\sim0.36\mu_B$
\cite{Cruz:08}), the mean distance between muon and iron moments in the
relevant phase is estimated to be $\sim0.5$ nm from an average of
$\delta_i$.  Given the unit cell size ($a=0.403$ nm, $c=0.874$ nm
\cite{R_Kamihara}), this would mean that more than a quarter of iron
atoms in the magnetic phase (i.e., $\simeq7$\% of the entire sample)
should serve as a source of local moments. It is unlikely that such a
significant fraction of iron atoms remains as impurities in the present
sample.

It might also be noteworthy that there is an anomaly near $T_{\rm
m2}\simeq12$ K in the susceptibility [the onset of ZFC/FC hysteresis in
Fig.~\ref{dcChi} and a steplike kink in Fig.~\ref{G_TFmulti}(a)].  This
seems to be in accordance with the onset of a steeper increase in
$\Lambda_{\rm m}$ below $T_{\rm m2}$, suggesting a change in magnetic
correlation.

$\mu$SR studies of LFAO-F have been made by a number of groups.
According to those preliminary studies, no clear sign of magnetism is
observed in the sample over relevant doping concentrations, except for a
weak relaxation observed far below $T_{\rm c}$ for $x=0.05$ and 0.075 in
ZF-$\mu$SR spectra and an unidentified additional relaxation observed in
TF-$\mu$SR spectra for
$x=0.075$.\cite{Carlo:08,Luetkens:08,Luetkens:08-2} This led us to
recall the sensitivity to chemical stoichiometry in the emergence of the
spin glass-like $A$-phase observed near the boundary between the
antiferromagnetic and superconducting phases in
CeCu$_2$Si$_2$.\cite{R_CeCu2Si2}.  In addition to the $A$-phase, the
present LFAO-F system exhibits a closer similarity to this classical
heavy-fermion superconductor such as the phase diagram against
pressure/doping.\cite{R_Monthoux} Further study of the dependence of
fractional yield for the magnetic phase with varying $x$ (in small steps
near the phase boundary) would provide further insight into the true nature
of these phases and the mechanism of superconductivity itself that is
working behind the coexistence/competition.

\par

In summary, it has been revealed by our $\mu$SR experiment that
superconducting and magnetic phases coexist in
LaFeAs(O$_{0.94}$F$_{0.06}$) with $x=0.06$. These two phases
simultaneously develop just below $T_{\rm c}$, strongly suggesting an
intimate and intrinsic relationship between these two phases.  The
result of TF-$\mu$SR measurement suggests that the superconductivity of
LaFeAs(O$_{0.94}$F$_{0.06}$) cannot be explained by the conventional
weak-BCS model (single gap, $s$-wave).

\par

We would like to thank the TRIUMF staff for their technical support
during the $\mu$SR experiment. This work was partially supported by the
KEK-MSL Inter-University Program for Oversea Muon Facilities and by a
Grant-in-Aid for Creative Scientific Research on Priority Areas from the
Ministry of Education, Culture, Sports, Science and Technology, Japan.


\end{document}